# Measurement of Conduction and Valence Bands g-factors in a Transition Metal Dichalcogenide Monolayer

C. Robert[1], H. Dery[2,3], L. Ren[1], D. van Tuan[2], E. Courtade[1], M. Yang[2], B. Urbaszek[1], D. Lagarde[1], K. Watanabe[4], T. Taniguchi[4], T. Amand[1] and, X. Marie[1]

[1]Université de Toulouse, INSA-CNRS-UPS, LPCNO, 135 Av. Rangueil, 31077 Toulouse, France
[2]Department of Electrical and Computer Engineering, University of Rochester, Rochester, New York 14627, USA
[3]Department of Physics, University of Rochester, Rochester, New York 14627, USA
[4]National Institute for Materials Science, Tsukuba, Ibaraki 305-004, Japan

*The electron valley and spin degree of freedom in monolayer transition-metal dichalcogenides can be manipulated in optical and transport measurements performed in magnetic fields. The key parameter for determining the Zeeman splitting, namely the separate contribution of the electron and hole g-factor, is inaccessible in most measurements. Here we present an original method that gives access to the respective contribution of the conduction and valence band to the measured Zeeman splitting. It exploits the optical selection rules of exciton complexes, in particular the ones involving inter-valley phonons, avoiding strong renormalization effects that compromise single particle g-factor determination in transport experiments. These studies yield a direct determination of single band g factors. We measure $g_{c1}= 0.86\pm0.1$, $g_{c2}=3.84\pm0.1$ for the bottom (top) conduction bands and $g_v=6.1\pm0.1$ for the valence band of monolayer $WSe_2$. These measurements are helpful for quantitative interpretation of optical and transport measurements performed in magnetic fields. In addition the measured g-factors are valuable input parameters for optimizing band structure calculations of these 2D materials.*

      The effective Landé *g*-factor of electrons, holes, and excitons in low-dimensional semiconductor systems has received considerable attention in the past 40 years since it provides precious information on the band structure [1–4]. The determination of the *g*-factor relies on measurement of the Zeeman energy splitting $\Delta E = g\mu_B B$, where $\mu_B$ is the electron's Bohr magneton and B is an external magnetic field which lifts the time inversion symmetry. Common techniques to measure the *g*-factor are electron spin resonance [5,6], Hanle effect [7], magneto-photoluminescence/absorption [8], spin quantum beats [9,10] or spin flip Raman scattering [11] experiments. It was first shown by Roth *et al.* that electrons in semiconductors can have an effective *g*-factor that differs substantially from the free-electron value $g_0=2$ as a consequence of the spin–orbit interaction (SOI), which couples the orbital motion with the spin degree of freedom [12].

Monolayer Transition-metal dichalcogenides (ML-TMDs) are ideal two-dimensional (2D) semiconductor systems characterized by large SOI with original optoelectronic and spin/valley properties [13–15]. Magneto- photoluminescence (PL) or reflectivity measurements with out-of-plane magnetic fields were performed on ML- $MoS_2$, $MoSe_2$, $MoTe_2$, $WS_2$ and $WSe_2$ [16–24]. These experimental investigations yield the exciton *g*-factors, but do not give the respective contribution linked to the conduction band (CB) and valence band (VB) *g*-factors.

The Zeeman splitting between right and left circularly-polarized light components $\sigma^+/\sigma^-$ (defined as $E_{\sigma^+} - E_{\sigma^-} = g\mu_B B$) yields a bright exciton *g*-factor close to $g \approx -4$ for most ML-TMDs. Surprisingly, this measured value is in agreement with a simple "atomic physics" model where the CB and VB *g*-factors result simply from the addition of three contributions, labelled by spin, valley and orbital terms [19,24,25]. However, the exciton *g*-factor with this approach just reflects the contribution of the VB orbital terms. As a consequence, no decisive information can be obtained on the CB or VB *g*-factor values. Moreover, this simple model usually fails to predict the carrier *g*-factors in other semiconductor structures (for instance the well-known *g*-factor of holes in GaAs [18]). It has indeed been shown that accurate determination of the *g*-factor requires precise description of the band-structure electronic states and in particular the mixing induced by the SOI [26,27]. In addition, the measurement of both CB and VB *g*-factors ($g_c$ and $g_v$) should provide valuable information on the electronic structure in ML-TMDs, for which many unknowns persist. For example, the value of the effective mass in the CB is still under debate [28–31]. The knowledge of the single particle *g*-factor is also essential to interpret the magneto-transport experiments in which the large carrier density induces strong renormalization effects due to many-body interactions [29,32].

In this Letter we present magneto-PL measurements performed on a charged adjustable ML-WSe$_2$ sketched in Fig. 1a. Details on the sample fabrication and experimental setup can be found in the Supplemental Material S1. We show that knowledge of the selection rules associated to optical transitions of different exciton complexes, in particular the dark positive trion and its zone-edge phonon replica, allows us to measure the *g*-factors of the bottom CB ($g_{c1}$) and top VB ($g_v$); see Fig. 2a. From the measured neutral exciton *g*-factor, we can then deduce the *g*-factor of the top CB ($g_{c2}$). We find $g_{c1}=0.86\pm0.1$, $g_{c2}=3.84\pm0.1$ and $g_v=6.1\pm0.1$ in ML-WSe$_2$. These values differ from the predictions based on simple additive contributions of the spin, valley and orbital components. However our measured CB and VB *g*-factors are in very good agreement with recent advanced Density Functional Theory (DFT) calculations taking into account the fine characteristics of the band structure [33–35]. The experimental technique presented here to determine the *g*-factors could be applied to other ML-TMDs in the future. Finally, we evidence a clear valley-dependent broadening of the dark positive trion PL lines, which results from Coulomb interaction between the bound trion complexes and the Fermi-sea.

**Experimental results**
We first present low temperature (T=5K) PL intensity and reflectivity contrast as a function of the hole doping density (tuned by the applied voltage V). The estimation of the carrier density is presented in the Supplemental Material S2. The neutrality region is easily identified in reflectivity when only the signature of the neutral exciton $X_0$ is seen (Fig. 1b). When we increase the hole density, a clear signature of positively charged exciton $X^+$ (formed by one electron and two holes of opposite spins) is observed in agreement with previous reports [36–40]. The method presented below to determine the CB and VB *g*-factors will be applied first in the very low doping density regime, typically p~$10^{11}$ cm$^{-2}$, in order to avoid band-gap renormalization effects [41]. Note that this doping density is two orders of magnitude weaker than the critical Mott density in ML-WSe$_2$ [42,43]. The PL intensity plot as a function of doping in Fig. 1c clearly evidences exciton complex transitions already identified in the literature: in addition to the neutral bright exciton ($X_0$) and the positive bright trion ($X^+$), we observe the peaks corresponding to the neutral dark (spin forbidden) exciton $X_D$, the positive dark trion ($X_D^+$) and its phonon replicas ($X_{D,K3}^+$) and ($X_{D,\Gamma5}^+$) [44–46], which lie 26 and 21 meV below ($X_D^+$), respectively. The notation of phonons K3 and $\Gamma_5$ has its origin in the Koster notation of the K-point irreducible representations of the $C_{3h}$ point double-group [47,48].

Figure 2b presents schematically the single-particle band structure of ML-WSe$_2$ at the vicinity of its CB and VB edges, along with optical selection rules associated to the radiative recombination of $X_D^+$ and $X_{D,K3}^+$ in an out-of-plane positive magnetic field (B>0). It will allow us to present the method which yields the determination of CB and VB g-factors.

Because of the interplay between SOI and the lack of inversion symmetry, absorption/emission of right and left circularly-polarized light occurs in the inequivalent valleys K+ and K- of the 2D hexagonal Brillouin zone [49–53]. The SOI yields a splitting of $\Delta_c$ ~25 meV and $\Delta_v$~450 meV between spin-up and down bands in the K+ valley (and opposite sign values in the K- valley). Here we will focus on the lowest energy transitions involving only the top VB, characterized by a g-factor $g_v$. The bottom and top CB g-factors are labelled $g_{c1}$ and $g_{c2}$, respectively. The g-factor of a given transition with in-plane dipole ($X_0$, $X_{D,K3}^+$) writes simply: $E_{\sigma+} - E_{\sigma-} = g\mu_B B$. For transitions with an out-of-plane dipole ($X_D$, $X_D^+$), the light polarization is perpendicular to the ML plane and we define the g-factor by $E_{K+} - E_{K-} = g\mu_B B$ (see Figs. 2b and 2c). For the well-known bright ($X_0$) and dark ($X_D$) neutral exciton transition, one can easily check that:

$$g_{X0}=-2(g_v-g_{c2})$$
$$g_{XD}=-2(g_v-g_{c1}). \tag{1}$$

Figure 3a shows the measured transition energies for a magnetic field varying between B=-9 and B=+ 9 Tesla. $X_0$ and $X_D$ are measured at charge neutrality, while $X_D^+$ and $X_{D,K3}^+$ are measured in a low p-doping regime ($p$=1.4 $10^{11}$ cm$^{-2}$). In agreement with previous reports, we find $g_{X0}$=-4.5±0.1 and $g_{XD}$=-10.2±0.1. As these two transitions imply two different CBs, these measurements cannot yield a determination of the CB and VB g-factors (see Eq. (1)). On the other hand, the selection rules associated with the positive dark trion optical transitions $X_D^+$ and $X_{D,K3}^+$ allow us to solve this problem. As illustrated in Figs. 2b and 2c, the positive dark trion optical transitions $X_D^+(K+)$ or $X_D^+(K-)$ denote recombination of the neutral dark exciton component and they occur between opposite VB and CB spins of the same valley (the second hole in the time-reversed valley can be considered as a "spectator"). As expected, the extracted g-factor of this transition is very close to the one of the neutral dark exciton since $g_{X_D^+}$ =-2($g_v$-$g_{c1}$); we measure $g_{X_D^+}$ =-10.5±0.1 [45,46].

In contrast, optical transitions that are associated with the K3 phonon replica, $X_{D,K3}^+$, involve the second hole of the trion, which is no more a "spectator" for the optical transition (Figs. 2b and 2c) [54]. As a consequence, the energy difference between the optical transitions $X_D^+$ and $X_{D,K3}^+$ will depend only on single band g-factors $g_{c1}$ and $g_v$ and the energy of the phonon $E_{K3}$ :

$$\Delta E_1 = E\ (X_D^+(K+))\ - E(\ X_{D,K3}^+(\sigma\text{-})\ ) = E_{K3} - 2g_v\mu_B B,$$
$$\Delta E_2 = E\ (X_D^+(K-))\ - E(\ X_{D,K3}^+(\sigma+)\ ) = E_{K3} + 2g_v\mu_B B,$$
$$\Delta E_3 = E\ (X_D^+(K-))\ - E(\ X_{D,K3}^+(\sigma\text{-})\ ) = E_{K3} - 2g_{c1}\mu_B B,$$
$$\Delta E_4 = E\ (X_D^+(K+))\ - E(\ X_{D,K3}^+(\sigma+)\ ) = E_{K3} + 2g_{c1}\mu_B B. \tag{2}$$

Figure 3b presents the variation of $\Delta E_2 - \Delta E_1$ as a function of the applied magnetic field. The slope of the curve ($4g_v\mu_B$) in Fig. 3b yields a direct determination of the valence band *g*-factor. We find $g_v$=6.1±0.1. Figure 3c presents the magnetic field variation of the difference $\Delta E_4 - \Delta E_3$ following the same procedure. The slope gives a direct determination of the bottom CB *g*-factor; we measure $g_{c1}$=0.86±0.1. Then, using the measured *g*-factor of the neutral exciton transition (Fig. 3a and Eq. (1)), we can deduce the top CB *g*-factor: $g_{c2}$=3.84±0.1. Finally, we extract the bottom VB *g*-factor, $g_{v2}$, using the relation $g_{X0}(B)=-2(g_{v2}-g_{c1})$, where $g_{X0}(B)$ corresponds to the type-B neutral exciton (optical transition with in-plane dipole between bottom VB and CB). From the previously measured value of $g_{X0}(B)$=-3.9±0.5 [24], and our result for $g_{c1}$=0.86±0.1, we get that $g_{v2}$=2.81±0.5.

**Discussion**

Zeeman splitting and corresponding *g*-factors were calculated in ML-TMDs using DFT, tight-binding or k.p approaches [33–35,55,56]. As shown in table I, our measurements are in excellent agreement with the single-band *g*-factors calculated recently by DFT. Despite the largest uncertainty associated to the measurement of the type-B exciton Zeeman energy that we took from Ref. [24], we note that the extracted value of the bottom VB *g*-factor ($g_{v2}$=2.81±0.5) is also close to the DFT calculated one (3.15) [33].

Moreover, our measurements clearly demonstrate that the simple model for calculating the *g*-factor based on the additive contribution of the magnetic coupling to the electron spin, valley and orbital angular momenta from the transition-metal atoms is oversimplified [19,25,57]. Though it gives a top valence band $g_v$ value close to the measured one (5.5), it fails to predict the *g*-factor of the CB as the simple calculation gives $g_{c2}$=3.5 and $g_{c1}$=1.5, assuming identical CB and VB mass ($0.4m_0$). This shows that in a similar way to other semiconductors, the calculation of the *g*-factor in ML-TMDs requires a rather detailed description of the band structure that takes into account subtle effects of the SOI [12,26,27,58,59].

Beyond the importance of the presented technique to extract the single particle *g*-factor in the CB and VB of ML-TMDs, the results of this work merit discussion of three important points.

The first one deals with the relation between the free electron (or hole) *g*-factor and that of excitons/trions. Our interpretation of the experiment assumes that the exciton Zeeman splitting is the sum of the Zeeman splitting energies in the CB and VB. Similar to the case of various semiconductors, we have neglected effects linked to the exciton wavefunction [2,58,59]. As the exciton binding energy in ML-TMDs is of the order of few hundreds meV (*i.e.* large extension in reciprocal space), one can question if the single-particle approach is accurate enough. Several calculations based on DFT coupled to Bethe-Salpeter Equation predicted a reduction of the exciton *g*-factor up to 30% compared to the single-particle approach [33,55], resulting from a decrease of the magnetic moment away from the band extrema. However, we believe that the excitonic correction is negligible in ML-WSe$_2$ for two reasons: (i) The measured values of *g*-factors match very well the predicted single band *g*-factors (table I) rather than the ones calculated with the exciton corrections. (ii) Previous measurements of the neutral exciton *g*-factor showed that $g_n$= -4.3 ± 0.2 for the ground and excited exciton states from *n*=1 to *n*=4 [60]. While these states are characterized by distinct extension in *k*-space, their *g*-factors are essentially the same within the experimental uncertainty.

The second point deals with the carrier-density dependence of the *g*-factors. It is well known that exchange interactions in quantum wells of III-V semiconductors lead to enhancement of the effective *g*-factors [61]. This enhancement has been recently evidenced in magneto-transport experiments of heavily doped ML-TMDs [32,62,63], and in Landau-quantized excitonic absorption spectroscopy of bright trions [36,64,65]. Wang *et al.* reported $g_v$ ~8.5 and $g_{c1}$~4.4 for a carrier density of ~6.10$^{12}$ cm$^{-2}$ in ML-WSe$_2$ [39], whereas Liu *et al.* found $g_v$ ~15 and $g_{c1}$~2.5 for densities larger than 10$^{12}$ cm$^{-2}$ [64]. In contrast, our measured values, $g_v$=6.1 and $g_{c1}$=0.86, show relatively little change when the gate-induced hole density changes from ~10$^{11}$ cm$^{-2}$ (Fig. 3) to 1.7x10$^{12}$ cm$^{-2}$ (see Supplementary Material S3), during which the corresponding Landau level filling factor at B=9T increases from $\nu < 1$ to $\nu > 6$. The reason for the disparity in the reported *g*-factor values remains an open question, and the relation between the energy-shift of various optical transitions and exchange interactions at large doping densities is yet to be quantified. Additional work is in progress and will be presented elsewhere.

Finally, we provide evidence that trions are not readily dissociated at elevated charge densities. Figures 4a and 4b show the magneto-PL spectra at B=9T, where holes primarily populate the K+ valley and their densities are 1.4x10$^{11}$ cm$^{-2}$ and 1.7x10$^{12}$ cm$^{-2}$, respectively. Inspecting the behavior of the positive dark trion with electron in K+, Fig. 4b shows that the zero-phonon optical transition, $X_D^+(K+)$, is largely quenched. This observation is part of a universal behavior seen in semiconductors wherein optical transitions of few-body bound complexes are broadened and eventually quenched if the spin, valley, and energy band of the recombining electron or hole are similar to those of the Fermi-sea particles (here referring to holes in the K+ valley). The point we wish to emphasize is that the quenched optical transition does not mean that the trion dissociates or that the trion picture is inadequate at elevated charge densities. This fact is vividly shown by comparing the optical transition, $X_{D,K3}^+(\sigma-)$, in Figs. 4a and 4b, corresponding to zone-edge phonon replica of the same trion. Not only that its peak amplitude is nearly 50% stronger in the higher density case, its full width at half maximum (FWHM) is narrower: ~1.7meV at 1.7x10$^{12}$ cm$^{-2}$ vs ~3.4meV at 1.4x10$^{11}$ cm$^{-2}$. We attribute this counterintuitive narrowing to the longer lifetime of the trion when its recombination channels through the zero-phonon and zone-center phonon replica ($\Gamma_5^-$) are quenched. Thus, while the quenched behavior of $X_D^+(K+)$ may suggest that the trion is no longer bound, its phonon replica refutes this possibility. This interpretation is further supported by the opposite behavior of the positive dark trion with electron in K- due to opposite labeling of the spectator and recombining holes. Here, the zero-phonon optical transition, $X_D^+(K-)$, increases in intensity when the hole density increases, whereas its zone-edge phonon replica, $X_{D,K3}^+(\sigma+)$, becomes weaker and its FWHM is broadened to ~5.3meV at 1.7x10$^{12}$ cm$^{-2}$. All in all, by comparing the optical transitions of the dark positive trion (zone-edge phonon replica vs the zero-phonon and/or zone-center phonon replica), we can conclude that the trions remain bound in the presence of charge carriers in ML-TMDs [66–71]. Rather than trion dissociation, the measured broadening/quenching seen in their optical transitions could be interpreted as suppressed recombination due to enhanced Coulomb scattering of the recombining hole (electron) when it has similar spin and valley quantum numbers to those of the Fermi-sea holes (electrons).

In summary, we have performed magneto-photoluminescence spectroscopy in a gated ML-WSe$_2$ device. Based on the knowledge of the optical selection rules of different exciton complexes, we have proposed a new method to measure the single particle *g*-factor. Our measurements should make it possible to improve the band structure calculations in monolayer transition-metal dichalcogenides, in particular the dispersion curves of the conduction bands which are still little known. Knowledge of the single band *g*-factors should be valuable for understanding the properties of van der Waals heterostructures in which interlayer or Moiré exciton transitions could be identified thanks to their Zeeman splitting. Finally, this work shows that by comparing optical transitions of the bare positive dark trion and its zone-edge phonon replica, one can better understand the interaction between bound trions and free charged carriers at elevated charge densities.

**Acknowledgments**
We thank Mathieu Pierre for his contribution to e-beam lithography. This work was supported by Agence Nationale de la Recherche funding ANR 2D-vdW-Spin, ANR VallEx and ANR MagicValley. The work at Rochester was funded by the Department of Energy, Basic Energy Sciences, under Contract No. DE-SC0014349. X.M. also acknowledges the Institut Universitaire de France.

|  | Measurements (this work) | Calculations Deilmann et al [33] | Calculations Woźniak et al [35] | Calculations Förste et al [34] | Calculations (spin/valley/orbital terms) [19,25] |
| --- | --- | --- | --- | --- | --- |
| $g_{c1}$ | **0.86** | **0.99** | **0.87** | **0.9** | **1.5** |
| $g_{c2}$ | **3.84** | **3.97** | **3.91** | **3.9** | **3.5** |
| $g_v$ | **6.1** | **5.91** | **5.81** | **5.9** | **5.5** |

Table I

# Figures

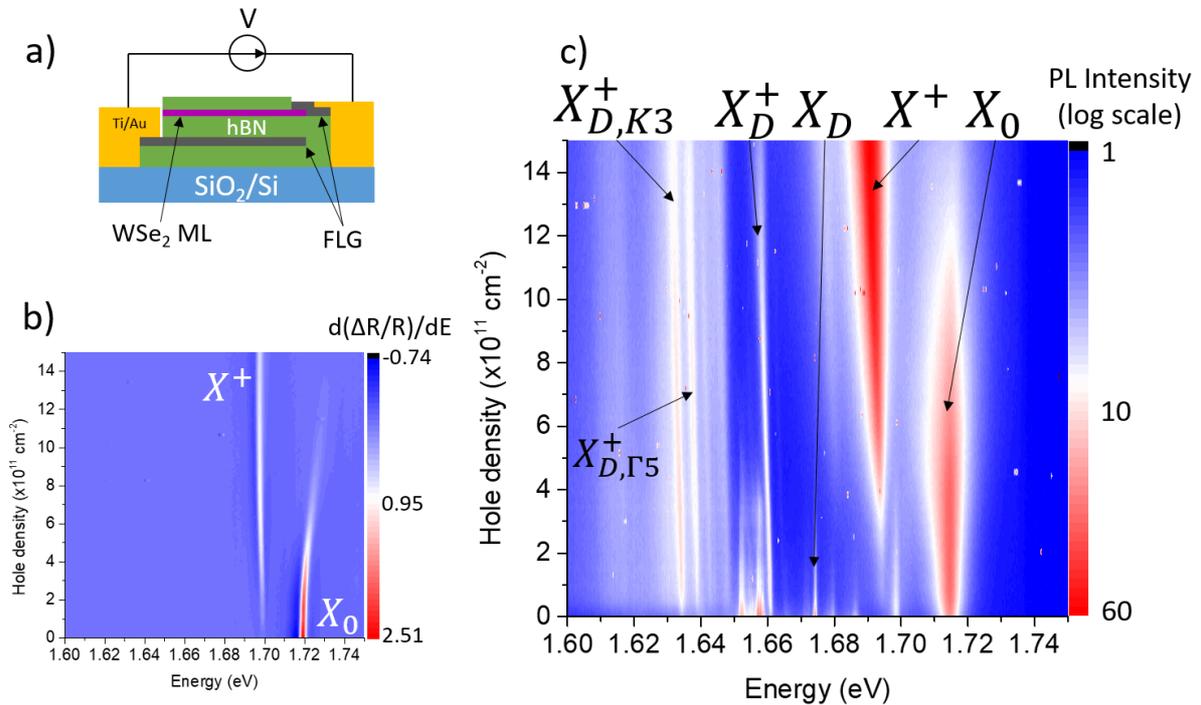

**Figure 1**. (a) Sketch of the sample. A first hBN flake is exfoliated onto the SiO$_2$(80nm)/Si substrate. Then few layers graphene (FLG) are deposited as a backgate. A second hBN flake of 140 nm is then transferred and act as the dielectric layer of the parallel plate capacitance. Then the ML-WSe$_2$ is transferred and is contacted with a second FLG flake before being capped by a thin top hBN. (b) First derivative of the reflectivity contrast and (c) photoluminescence intensity as a function of hole doping for a magnetic field B=0T.

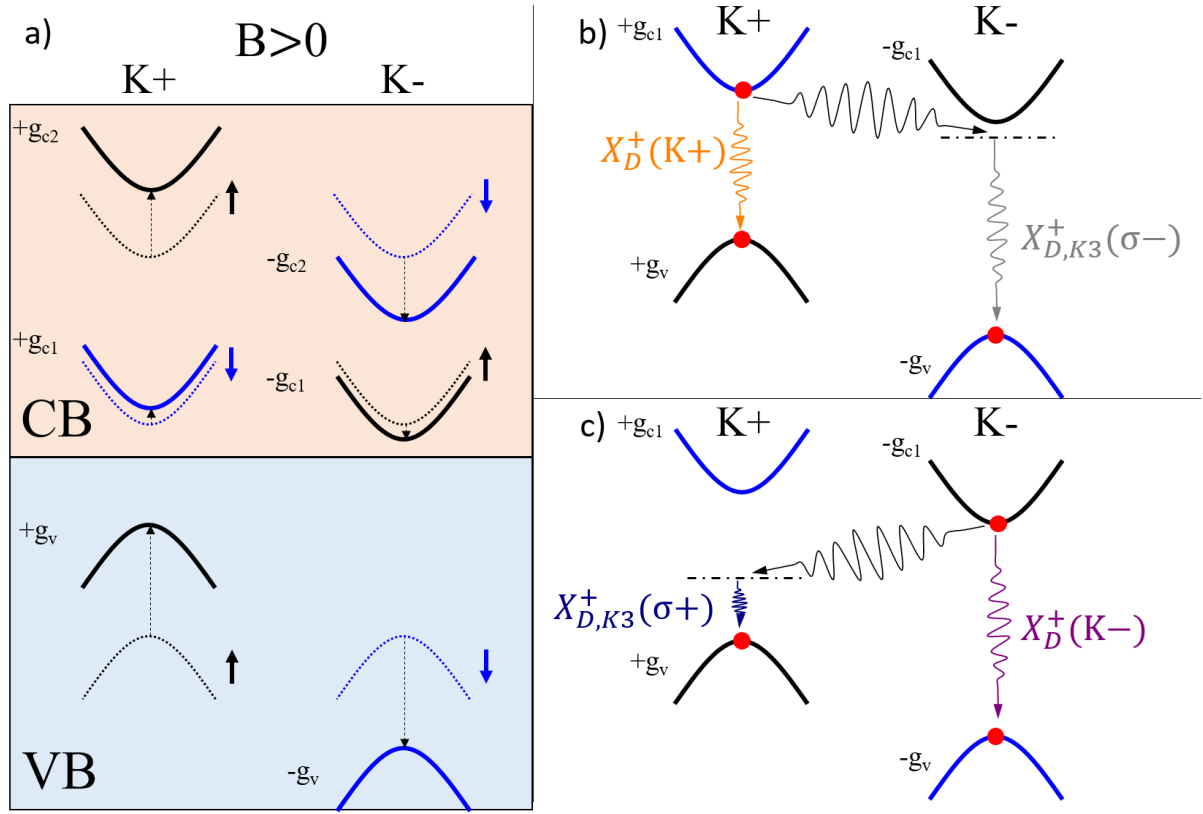

**Figure 2**. (a) Schematics of the band structure in valleys K+ and K- of ML-WSe$_2$ at the neutrality point; the dotted and full lines correspond to bands with B=0 and B>0, respectively. The vertical dotted lines indicate the magnetic field induced shift of the bands. Only the top valence band is considered here (type-A optical transitions). Schematics of the band structure for B>0 and hole doping displaying the optical transition corresponding to the K$_3$ phonon replica of positive dark trion, $X_{D,K3}^+$, in (b) valley K- and (c) valley K+. The optical transition of $X_D^+$ is also indicated. The top CBs are not displayed.

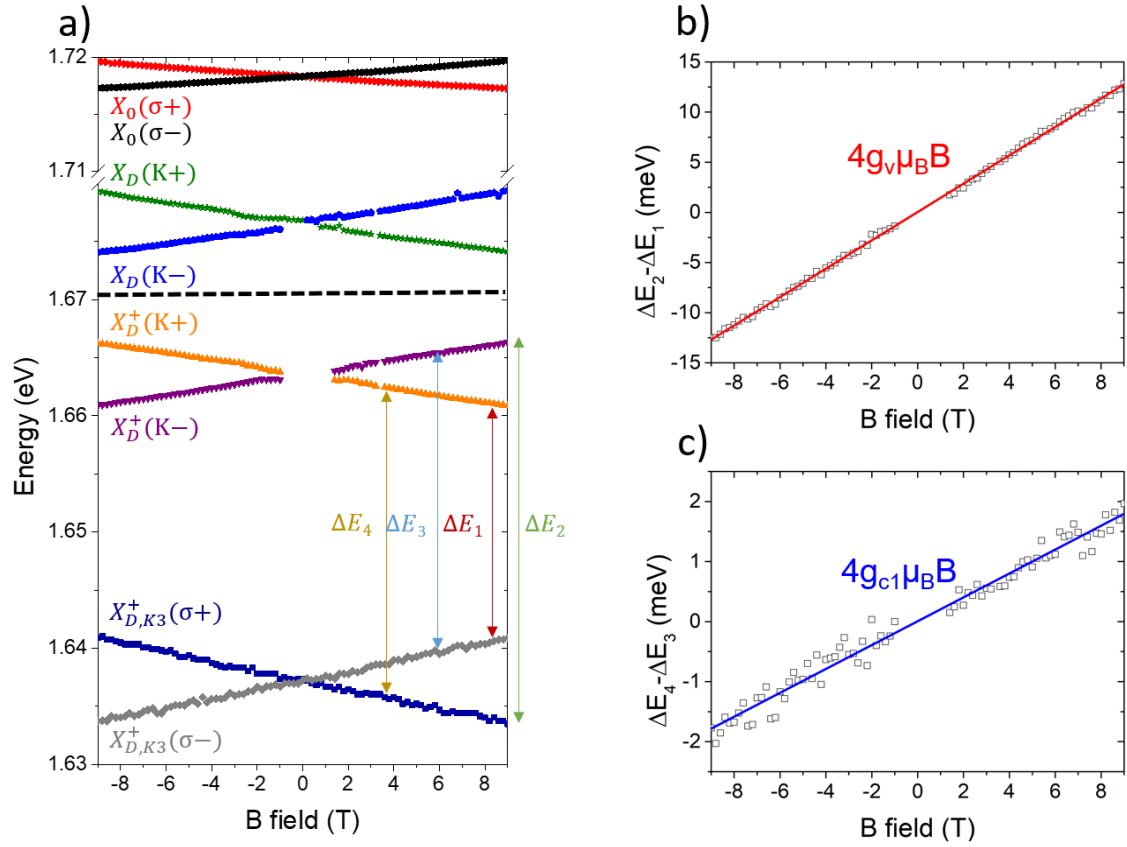

**Figure 3.** (a) Magnetic field dependence of bright and dark excitons at the neutrality point of the device and positive dark trion, $X_D^+$, and its phonon replica involving the inter-valley phonon K$_3$, $X_{D,K3}^+$, measured for $p=1.4 \; 10^{11}$ cm$^{-2}$.

Magnetic field dependence of the energy difference of the optical transitions (b) $\Delta E_2 - \Delta E_1$ and (c) $\Delta E_4 - \Delta E_3$ yielding the determination of $g_v$ and $g_{c1}$ (see arrows in (a)).

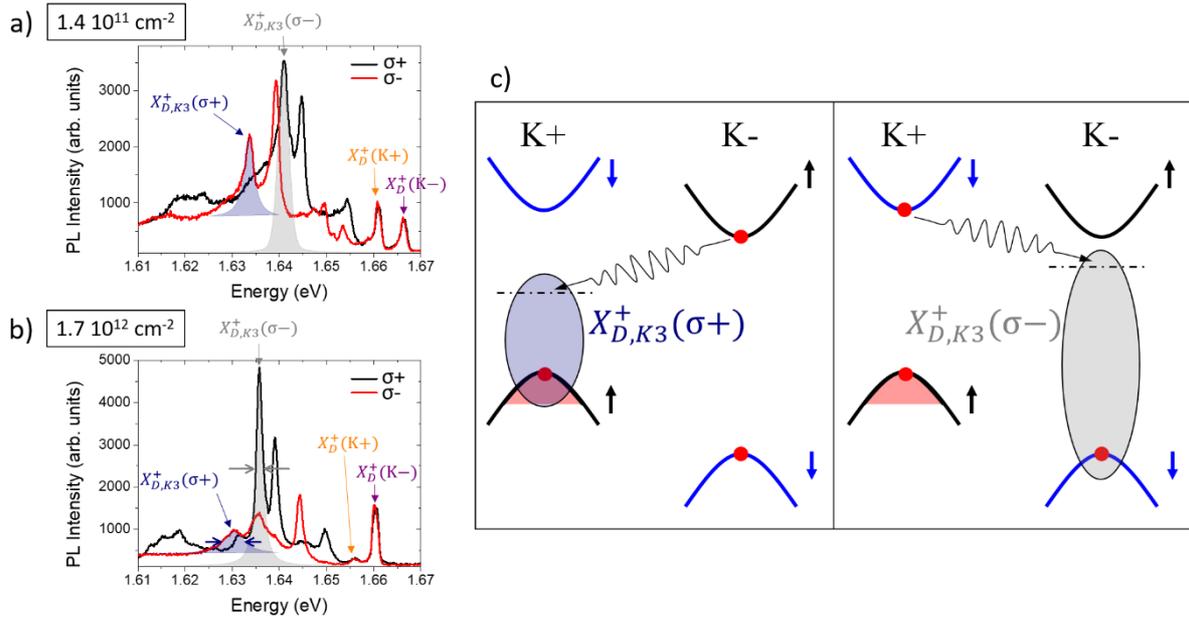

**Figure 4.** Photoluminescence spectra measured at B=9 T for (a) low (1.4 $10^{11}$ cm$^{-2}$) and (b) large (~1.7 $10^{12}$ cm$^{-2}$) hole doping densities; (c) schematics of the optical transition associated to the K$_3$ phonon replica of the positive dark trion, $X^+_{D,K3}$.

# Supplemental Material for "Measurement of Conduction and Valence Bands g-factors in a Transition Metal Dichalcogenide Monolayer"

C. Robert[1], H. Dery[2,3], L. Ren[1], D. van Tuan[2], E. Courtade[1], M. Yang[2], B. Urbaszek[1], D. Lagarde[1], K. Watanabe[4], T. Taniguchi[4], T. Amand[1] and, X. Marie[1]

[1]Université de Toulouse, INSA-CNRS-UPS, LPCNO, 135 Av. Rangueil, 31077 Toulouse, France
[2]Department of Electrical and Computer Engineering, University of Rochester, Rochester, New York 14627, USA
[3]Department of Physics, University of Rochester, Rochester, New York 14627, USA
[4]National Institute for Materials Science, Tsukuba, Ibaraki 305-004, Japan

## S1. Experimental set-up and sample

We have fabricated a van der Waals heterostructure made of an exfoliated ML-WSe$_2$ embedded in high quality hBN crystals [1] using a dry stamping technique [2]. The structure is sketched in Fig. 1a of the main text. We use few layers of graphene for the back gate and to contact the ML-WSe$_2$. Following the heterostructure fabrication, we performed e-beam lithography and Ti/Au deposition on the few layer graphene flakes to electrically contact the structure.

Magneto-PL experiments at T = 5 K and in magnetic fields up to 9 T have been carried out in an ultra-stable confocal microscope [3]. The detection spot diameter is about 700 nm. The sample is excited by a He-Ne laser (1.96 eV) with linear polarization and both circular σ+ and σ- polarized PL signals are detected using a liquid crystal retarder and an analyzer in the detection path. The average laser power is around 10 µW, which is still in the linear absorption regime. The PL emission is dispersed in a monochromator and detected with a Si-CCD camera.

## S2. Evaluation of the carrier density

The carrier density can be estimated using a simple plate capacitance model knowing the applied voltage (V), the hBN thickness $t$ (140 nm in our device) and using a hBN dielectric constant of $\varepsilon_{hBN}$ ~3 [4,5]. The change of hole density $\Delta p$ is related to a change of bias voltage $\Delta V$ by $\Delta p = \frac{\varepsilon_0 \varepsilon_{hBN}}{e*t} \Delta V$.

Alternatively, we can use the oscillations in the reflectivity spectrum of the bright exciton as a function of gate voltage observed at +9 T (see Figure S1(b)). As demonstrated in Ref [6], these oscillations are due to the interaction of the exciton with the quantized Landau levels of the hole Fermi sea (see the sketch of Figure S1(a)). The period of the oscillations $\Delta V_{LL}$ is related to the filling of one Landau level $P_{LL} = \frac{eB}{2\pi\hbar}$ =2.18 10$^{11}$ cm$^{-2}$. We can thus calculate the hole density as a function of the gate voltage by: $\Delta p = \Delta V \frac{P_{LL}}{\Delta V_{LL}}$. This yields the same estimation of the carrier density as the one deduced from the capacitance model. The advantage of this method is that it does not require knowledge of material parameters.

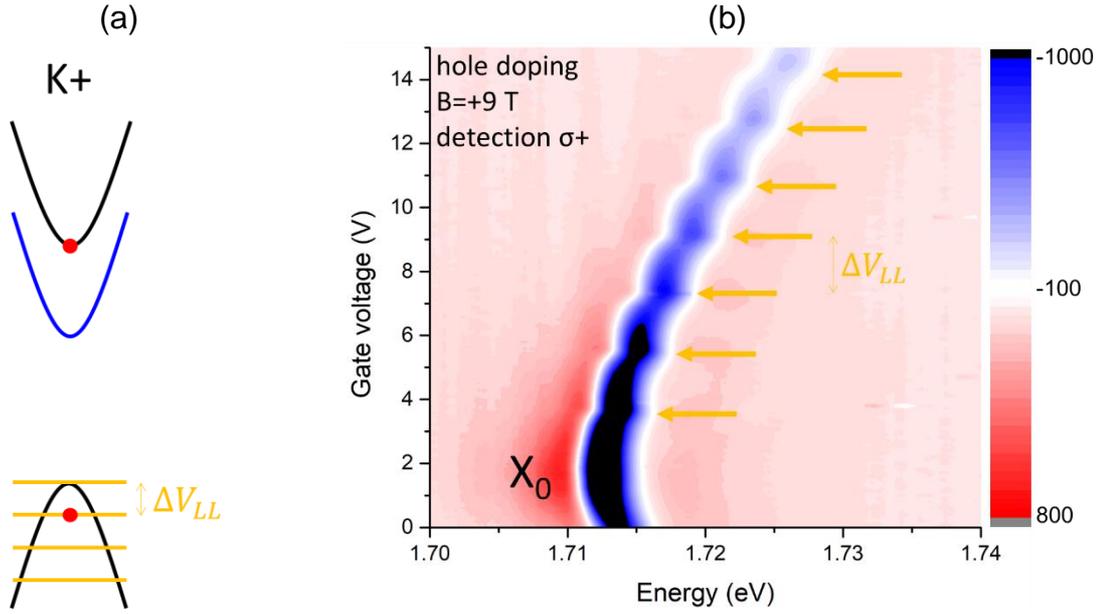

*Figure S1: (a) Sketch of the band structure of the K+ valley at positive magnetic field showing the Landau levels. (b) First derivative of the reflectivity contrast with σ+ detection as a function of the gate voltage at +9 T.*

## S3. Variation of the g factor with increasing the hole density

The determination of the Landé g-factors presented in the main text has been performed at small hole density ($1.4 \times 10^{11}$ cm$^{-2}$) to avoid many body effects. We show in Figure S2 the variations of the valence band g-factor using the same method when we increase the hole density.

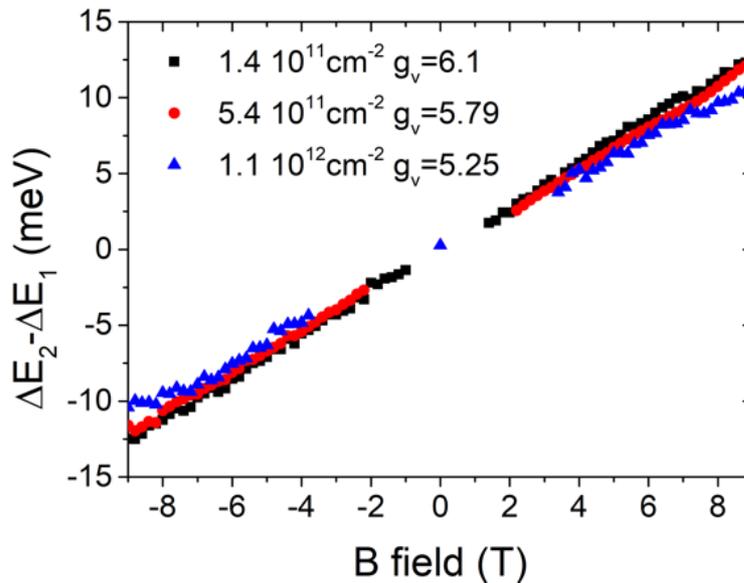

*Figure S2: $\Delta E_2 - \Delta E_1$ as a function of the magnetic field yielding the determination of $g_v$ for three values of hole density.*